\documentclass[aps,showpacs,preprintnumbers,amsmath,amssymb]{revtex4}
\UseRawInputEncoding
\usepackage{dcolumn}
\usepackage[dvips]{epsfig}
\usepackage{float}
\usepackage{graphics,epsfig}

\usepackage{graphicx}
\usepackage{epstopdf}
\usepackage{booktabs}
\usepackage{multirow}
\usepackage{threeparttable}
\usepackage{dcolumn}
\usepackage{bm}
\usepackage{gensymb}
\usepackage{footnote}
\usepackage{booktabs}
\usepackage{color}
\usepackage{pdfpages}

\begin{document}
\baselineskip=0.8 cm
\title{Polarized image of an equatorial emitting ring around a 4D Gauss-Bonnet black hole}

\author{Xin Qin$^{1}$,
Songbai Chen$^{1,2}$\footnote{Corresponding author: csb3752@hunnu.edu.cn},
Jiliang Jing$^{1,2}$ \footnote{jljing@hunnu.edu.cn}}
\affiliation{ $ ^1$ Department of Physics, Synergetic Innovation Center for Quantum Effects and Applications, Hunan
Normal University,  Changsha, Hunan 410081, People's Republic of China
\\
$ ^2$Center for Gravitation and Cosmology, College of Physical Science and Technology, Yangzhou University, Yangzhou 225009, People's Republic of China}

\begin{abstract}
\baselineskip=0.6 cm
\begin{center}
{\bf Abstract}
\end{center}

We have studied the polarized image of an equatorial emitting ring around a 4D Gauss-Bonnet black hole. Our results show that the effects of Gauss-Bonnet parameter on the polarized image depend on the magnetic field configuration, the observation inclination angle, and the fluid velocity. As the magnetic field lies in the equatorial plane, the observed polarization intensity increases monotonously with Gauss-Bonnet parameter in the low inclination angle case, and its monotonicity disappears in the case with high inclination angle.  However, as the magnetic field is vertical to the equatorial plane, the polarization intensity is a monotonously increasing function of Gauss-Bonnet parameter in the high inclination angle case. The changes of the electric vector position angle with Gauss-Bonnet parameter in both cases are more complicated. We also probe the effects of Gauss-Bonnet parameter on the Strokes $Q$-$U$ loops.
\end{abstract}

\pacs{ 04.70.Dy, 95.30.Sf, 97.60.Lf } \maketitle
\newpage
\section{Introduction}

The Event Horizon Telescope (EHT) collaboration \cite{EHT1,EHT2,EHT3,EHT4,EHT5,EHT6} recently released the first image of the supermassive black hole M87*, which is one of the most exciting events in physics. It provides evidence of the photon ring and indicates that the observational black hole astronomy has been entered a new era of rapid progress. In the first polarized images of the black hole M87* \cite{EHT7,EHT8}, a bright ring of emission with twisting
polarizations pattern brings us a lot of information about the electromagnetic radiation near black hole. Analyzing the corresponding polarization patterns could help us to probe the magnetic field configuration and the accretion
process around black hole, which is beneficial to gain an insight into physics in the strong field region near black holes.
Thus, a lot of effort have been focused on the study of polarized images of black holes \cite{PZJG1,PZJG2,PZJG3,PZJG4,PZJG5,PZJG6,PZJG7,PZJG8,PZJG9,PZJG10,PZJG11,PZJG12,Bel,QU1,QU2}.

In general, one must resort to numerical simulations to get an exact description of polarized image of emitter
around black hole, which is computationally expensive. Recently, a simple model has been used to investigate
the polarized images of axisymmetric fluids orbiting Schwarzschild and Kerr black holes arising from synchrotron emission in various
magnetic fields \cite{PZ1,PZ2}.
It is shown that the polarization signatures of the image, including linear polarization angle, relative polarized intensity and  strokes $Q-U$ loops, are dominated by black hole geometry together with the fluid velocity, magnetic field configuration, and observer inclination. Though only the emission from a single radius is considered in this model, the image of a finite thin disk can be produced by simply summing contributions from individual radii \cite{ringpz2}.

It is well known that the polarized patterns of black hole image depend on the strong gravity field near the black hole. Comparing the theoretical polarized patterns with the observed polarization signatures could help gain some characteristic information about the gravity field in the vicinity of black hole, and provide a potential tool to examine various theories of gravity including general relativity. Therefore, it is necessary to study polarized images of electromagnetic emission around black holes in various theories of gravity. Gauss-Bonnet gravity is a theory with higher curvature correction and its correction is precisely given by a combination of quadratic curvature terms, i.e., the so-called Gauss-Bonnet term. In four dimensional spacetimes, Gauss-Bonnet term is generally a total derivative and does not contribute to the equations of motion. However,
 Glavan et al. \cite{BH} recently find that some finite nontrivial effects of Gauss-Bonnet contributions can emerge in four dimensional spacetimes through multiplying the higher dimensional Gauss-Bonnet term by the factor $1/(D-4)$. Some new spherically symmetric black hole solutions are obtained in such 4D Einstein-Gauss-Bonnet gravity.
Although there are some criticisms on this naive 4D theory of Gravity and its the regularization scheme \cite{criticism1,criticism2,criticism3}, a well-defined theory is proposed to serve as a consistent realization of the 4D Einstein-Gauss-Bonnet
gravity where there exist two dynamical degrees of freedom with breaking the temporal diffeomorphism invariance \cite{ADM}. It is fortunate that the black hole solutions obtained by the naive regularization \cite{BH} are confirmed as exact solutions in this full theory \cite{ADM}, which has greatly contributed to the study of 4D Einstein-Gauss-Bonnet gravity on various aspects \cite{res1,res2,res3,res4,res5,res6,res7,res8,res9,res10,res11,res12,res13,res14,res15,res16,res17,res18}. In this paper, we will study the polarization image of an equatorial synchrotron emitting ring around the 4D Gauss-Bonnet black hole  \cite{BH} and to see the effects of Gauss-Bonnet parameter on the polarization image.

The paper is organized as follows: In Sec. II, we introduce briefly the 4D Gauss-Bonnet black hole \cite{BH} and review the calculation of the observed polarization image for the orbiting fluid model \cite{PZ1,PZ2}.
In Sec. III, We study the effects of the Gauss-Bonnet parameter on the black hole polarization image. Finally, we end the paper with a summary.

\section{Observed polarization field for the orbiting fluid model in the 4D Gauss-Bonnet black hole spacetime}

The 4D Gauss-Bonnet black hole solution is obtained firstly in \cite{BH}. Starting from the usual $D-$dimensional Gauss-Bonnet action  in a curved spacetime
\begin{equation}\label{Action}
S=-\frac{1}{16\pi{G}}\int{d^Dx}\sqrt{-g}\left[R+\alpha\left(R_{\mu\nu\rho\sigma}R^{\mu\nu\rho\sigma}-4R_{\mu\nu}R^{\mu\nu}+R^2\right)\right],
\end{equation}
and rescaling the Gauss-Bonnet coupling constant $\alpha\rightarrow\frac{\alpha}{D-4}$,
they took the limit $D\rightarrow4$ and finally obtained a neutral 4D Gauss-Bonnet black hole solution with the metric form
\begin{equation}\label{metric}
ds^2=-f(r)dt^2+\frac{1}{f(r)}dr^2+r^2d\theta^2+r^2\sin^2{\theta}d\psi^2,
\end{equation}
where
\begin{equation}\label{fr}
f(r)=1+\frac{r^2}{2\alpha}\left(1\pm\sqrt{1+\frac{8\alpha{M}}{r^3}}\right).
\end{equation}
Here $M$ is black hole mass and $\alpha$ is the Gauss-Bonnet coupling constant with dimension of length-squared.
Clearly, there are two different branches of the sign $\pm$ for the solution (\ref{metric}).  Here we consider only the negative branch because that it can asymptotically go over to the Schwarzschild black hole. Solving the equation $f(r)=0$, we can obtain its two roots
\begin{equation}\label{event horizon}
r_{\pm}=M\pm\sqrt{M^2-\alpha},
\end{equation}
which correspond to the event horizon radius and the Cauchy horizon radius, respectively.

\begin{figure}
\includegraphics[width=7cm ]{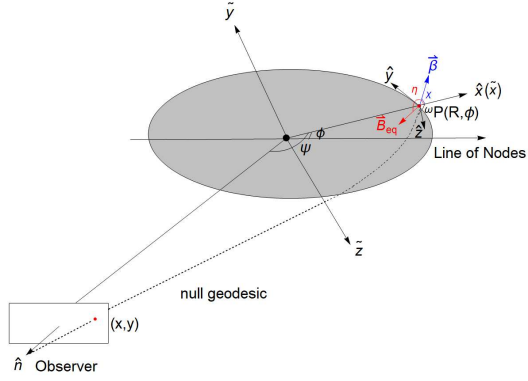}
\caption{Geometry of the orbital fluid model. There are two Cartesian coordinate systems at the center of the black hole and at point $P$, respectively. Two coordinate systems share the $\hat{x}$ ($\tilde{x}$)-axis along the radial direction from the center of the black hole to the radiation point $P$. The $\hat{y}$-axis in the local Cartesian coordinate system is the azimuthal direction. $B_{eq}$ and $\vec{\beta}$ denote the equatorial magnetic field and the fluid velocity, respectively.}
\label{tu2}
\end{figure}

As in refs.\cite{PZ1,PZ2}, we consider a synchrotron emitting ring around a 4D Gauss-Bonnet black hole (\ref{metric}) and the ring is assumed to lie in the equatorial plane of the black hole. The distant observer lies in the face-on orientation with an angle $\theta$ tilted from the normal direction of the emitting ring as shown in Fig.(\ref{tu2}). We consider a fluid element $P$ located at azimuthal angle $\phi$ measured from the line-of-nodes between the ring plane and the observer's sky plane.
The photon radiating from the point $P$ moves along the geodesic to the observer at infinity and it sweeps through the angle $\psi$ in the geodesic plane in the meantime. From the geometry in Fig.(\ref{tu2}),  one can find that the angle $\psi$ and the azimuth angle $\phi$ satisfy the relation
\begin{equation}
\cos\psi=-\sin\theta\sin\phi.
\end{equation}
In the 4D Gauss-Bonnet black hole spacetime (\ref{metric}), the angle $\psi$ swept by photon along the geodesic from the emitting ring ( with radius $R$ ) to observer at infinity can be expressed as
\begin{equation}\label{geodesics1}
\psi=\int^\infty_R\frac{dr}{r^2\sqrt{\frac{1}{b^2}-\frac{f(r)}{r^2}}},
\end{equation}
where $b$ is the impact parameter. Moreover, the  emission angle $\omega$ (between the emitted photon and the local
radial direction) is related to the four velocity $u^{\mu}$ by $\omega=\arctan(\frac{\sqrt{u^\psi{u_\psi}}}{u^r{u_r}})|_{r=R}$, which means
\begin{equation}\label{geodesics2}
\sin\omega=\frac{b\sqrt{f(R)}}{R}.
\end{equation}
With the above equations (\ref{geodesics1}) and (\ref{geodesics2}), we can get the relationship between $\psi$ and $\omega$ by numerical integration \cite{Bel}. Here we set photon energy measured by an observer at infinity is $k_t=-1$.  Thus, the corresponding orthogonal time component and orthogonal space component of photon's momentum at the location $P$ in the G-frame (the geodesic frame \cite{PZ1,PZ2}) can be expressed as
\begin{eqnarray}\label{energy conservation1}
k^{\tilde{t}}_{(G)}=\frac{1}{\sqrt{f(R)}},  \quad\quad \quad k^{\tilde{x}}_{(G)}=k^{\tilde{t}}_{(G)}\cos\omega,
 \quad\quad \quad k^{\tilde{y}}_{(G)}=0, \quad\quad\quad k^{\tilde{z}}_{(G)}=k^{\tilde{t}}_{(G)}\sin\omega,
\end{eqnarray}
where $k^{\tilde{y}}_{(G)} =0$ because the geodesic lies in the $\tilde{x}\tilde{z}$-plane shown in Fig.(\ref{tu2}). At the point $P$, we can build another local Cartesian frame called P-frame which shares a same $\hat{x}$-axis, but $\hat{y}$ axis is along in the azimuthal direction parallel to $\hat{\phi}$, and $\hat{z}$ perpendicular to the disk plane.
One can transform from one to the other by rotating  an angle $\xi$
around the $\tilde{x}$(or $\hat{x}$) axis. With the help of the unit vector
$\tilde{n}$ from the black hole $O$ towards the observer, one can find that the angle $\xi$ satisfies
\begin{equation}
\cos\xi=\frac{\cos\theta}{\sin\psi}, \quad\quad \sin\xi=\frac{\sin\theta\cos\phi}{\sin\psi}.
\end{equation}
Therefore, one can obtain the orthogonal components $k^{\hat{\mu}}$ in the P-frame \cite{PZ1,PZ2}
\begin{eqnarray}\label{energy conservation2}
k^{\hat{t}}_{(P)}=\frac{1}{\sqrt{f(R)}}, \quad\quad\quad k^{\hat{x}}_{(P)}=\frac{\cos\omega}{\sqrt{f(R)}},
\quad\quad\quad k^{\hat{y}}_{(P)}=-\frac{\sin\xi\sin\omega}{\sqrt{f(R)}},\quad\quad \quad k^{\hat{z}}_{(P)}=\frac{\cos\xi\sin\omega}{\sqrt{f(R)}}.
\end{eqnarray}
Supposing that the fluid at the point $P$ has a velocity $\vec{\beta}$ with angle $\chi$ from the $\hat{x}$-axis in the local P-frame
\begin{equation}
\vec{\beta}=\beta(\cos\chi\hat{x}+\sin\chi\hat{y}),
\end{equation}
one can obtain the orthogonal components of $k^{\hat{\mu}}_{(F)}$ in the fluid frame (F-frame) through a Lorentz boost  \cite{PZ1,PZ2}
\begin{eqnarray}\label{transform1}
&&k^{\hat{t}}_{(F)}=\gamma{k^{\hat{t}}_{(P)}}-\gamma\beta\cos\chi{k^{\hat{x}}_{(P)}}-\gamma\beta\sin\chi{k^{\hat{y}}_{(P)}}, \\ \nonumber
&&k^{\hat{x}}_{(F)}=-\gamma\beta\cos\chi{k^{\hat{t}}_{(P)}}+\left(1+(\gamma-1)\cos^2{\chi}\right){k^{\hat{x}}_{(P)}}+(\gamma-1)\cos\chi\sin\chi{k^{\hat{y}}_{(P)}}, \\ \nonumber
&&k^{\hat{y}}_{(F)}=-\gamma\beta\sin\chi{k^{\hat{t}}_{(P)}}+(\gamma-1)\cos\chi\sin\chi{k^{\hat{x}}_{(P)}}+\left(1+(\gamma-1)\sin^2{\chi}\right){k^{\hat{y}}_{(P)}}, \\ \nonumber
&&k^{\hat{z}}_{(F)}=k^{\hat{z}}_{(P)},
\end{eqnarray}
where $\gamma$ is the Lorentz factor, and $\gamma=\frac{1}{\sqrt{1-\beta^2}}$. The radiation emitted along $k^{\hat{\mu}}_{(F)}$
in the F-frame is Doppler shifted by the time it reaches the observer. The Doppler factor $\delta=\frac{k^{\hat{t}}}{k^{\hat{t}}_{(F)}}$
describes both gravitational redshift and Doppler shift from velocity.
The magnetic field in the F-frame is assumed to have a form
\begin{equation}
\vec{B}=B_r\hat{x}+B_\phi\hat{y}+B_z\hat{z},\quad \vec{B}_{eq}\equiv B_r\hat{x}+B_\phi\hat{y},
\end{equation}
and the angle $\zeta$ between the magnetic field $\vec{B}$ and the 3-vector $\vec{k}_{(F)}$ obeys \cite{PZ1,PZ2}
\begin{equation}
\sin\zeta=\frac{|\vec{k}_{(F)}\times\vec{B}|}{|\vec{k}_{(F)}||\vec{B}|}.
\end{equation}
The above factor plays an important role in the intensity of synchrotron radiation.
It is well known that the direction of the electric vector of light is along the vector $\vec{k}_{(F)}\times\vec{B}$. Thus, the 4 polarization vector $f^{\mu}_{(F)}$ can be expressed as
\begin{equation}
\vec{f}^t_{(F)}=0,\quad\quad\quad\quad \vec{f}^{\;i}_{(F)}=\bigg(\frac{\vec{k}_{(F)}\times\vec{B}}{|\vec{k}_{(F)}|}\bigg)^i,
\end{equation}
where $\vec{f}^t_{(F)}=0$ and $i=x, y, z$. and then the normalized polarization vector satisfies
\begin{equation}
\vec{f}^\mu_{(F)}\vec{f}_{\mu(F)}=\sin^{2}\zeta|\vec{B}|^{2}.
\end{equation}
Making use of the inverse Lorentz transform, one can obtain the components of the polarization vector $f^{\mu}$ of photon at the point $P$ in the P-Frame \cite{PZ1,PZ2}
\begin{eqnarray}\label{transform2}
%\nonumber to remove numbering (before each equation)
&&f^{\hat{t}}_{(P)}=\gamma{f^{\hat{t}}_{(F)}}+\gamma\beta\cos\chi{f^{\hat{x}}_{(F)}}+\gamma\beta\sin\chi{f^{\hat{y}}_{(F)}}, \\ \nonumber
&&f^{\hat{x}}_{(P)}=\gamma\beta\cos\chi{f^{\hat{t}}_{(F)}}+\left(1+(\gamma-1)\cos^2{\chi}\right){f^{\hat{x}}_{(F)}}+(\gamma-1)\cos\chi\sin\chi{f^{\hat{y}}_{(F)}}, \\ \nonumber
&&f^{\hat{y}}_{(P)}=\gamma\beta\sin\chi{k^{\hat{t}}_{(F)}}+(\gamma-1)\cos\chi\sin\chi{f^{\hat{x}}_{(F)}}+\left(1+(\gamma-1)\sin^2{\chi}\right){f^{\hat{y}}_{(F)}}, \\ \nonumber
&&f^{\hat{z}}_{(P)}=f^{\hat{z}}_{(F)}.
\end{eqnarray}
In the 4D Gauss-Bonnet black hole spacetime $(\ref{metric})$, the celestial coordinates $(x,y)$ for the photon
moving from point $P$ along the null geodesic to the observer at infinity are \cite{CB}
\begin{eqnarray}
&&x=-\frac{Rk^{\hat{y}}}{\sin\theta}, \\ \nonumber
&&y=R\sqrt{\left(k^{\hat{z}}\right)^2-\cot^2\theta\left(k^{\hat{y}}\right)^2}sgn(\sin\phi)
\end{eqnarray}
With the conserved Penrose-Walker constant $\kappa$ \cite{PWconstant}, one can easily calculate the polarized vector at the observer because both the real and imaginary parts of $\kappa$ are conserved along the null geodesic. At the point $P$ in the fluid, we have
\begin{eqnarray}\label{kappa0}
&&\kappa=\kappa_1+i\kappa_2, \quad\quad\quad
\kappa_1=\Psi_2^{-1/3}\left(k^{t}f^{x}-k^{x}f^{t}\right), \quad\quad\quad \kappa_2=\Psi_2^{-1/3}\left(k^{y}f^{z}-k^{z}f^{y}\right),
\end{eqnarray}
with
\begin{eqnarray}
&&k^{t}=\frac{1}{f(R)}, \quad\quad\quad\quad k^{x}=\sqrt{f(R)}k^{\hat{x}}_{(P)}, \quad\quad\quad\quad
k^{y}=\frac{k^{\hat{y}}_{(P)}}{R},\quad\quad\quad k^{z}=\frac{k^{\hat{z}}_{(P)}}{R}, \\ \nonumber
&&f^{t}=\frac{f^{\hat{t}}_{(P)}}{\sqrt{f(R)}}, \quad\quad\quad f^{x}=\sqrt{f(R)}f^{\hat{x}}_{(P)}, \quad\quad\quad\quad
f^{y}=\frac{f^{\hat{y}}_{(P)}}{R},\quad\quad\quad f^{z}=\frac{f^{\hat{z}}_{(P)}}{R}.
\end{eqnarray}
Here $\Psi_2$ is the Weyl scalar with the form
\begin{equation}
\Psi_2=-\frac{M(R^3+2M\alpha)}{R^{3/2}(R^3+8M\alpha)^{3/2}}.
\end{equation}
Thus, the normalized polarization electric field vector $\vec{E}$ along the $x$  and  $y$ directions in the observer's sky can be given by \cite{Himwich}
\begin{eqnarray}\label{EV0}
&&E_{x,norm}=\frac{y\kappa_2+x\kappa_1}{\sqrt{\left(\kappa_1^2+\kappa_2^2\right)\left(x^2+y^2\right)}}, \\ \nonumber
&&E_{y,norm}=\frac{y\kappa_1-x\kappa_2}{\sqrt{\left(\kappa_1^2+\kappa_2^2\right)\left(x^2+y^2\right)}}, \\ \nonumber
&&E_{x,norm}^2+E_{y,norm}^2=1.
\end{eqnarray}
For the synchrotron radiation, the intensity
of linearly polarized light that reaches the
observer from the source point $P$ can be approximated as \cite{PZ1,PZ2}
\begin{equation}\label{expression of intensity1}
|I|\sim\delta^{3+\alpha_\nu}l_p|\vec{B}|^{1+\alpha_\nu}\sin^{1+\alpha_\nu}\zeta,
\end{equation}
where the power $\alpha_\nu$ depends on the ratio of the emitted photon energy $h\nu$ to the disk temperature $kT$. The quantity $l_p$ denotes the geodesic path length for the photon traveling through the emitting region, which is can be expressed as
\begin{equation}
\quad l_p=\frac{k^{\hat{t}}_{(F)}}{k^{\hat{z}}_{(F)}}H.
\end{equation}
$H$ is the height of the disk which can be taken to be a constant for simplicity. As in refs.\cite{PZ1,PZ2}, we set $\alpha_\nu=1$, and then
the observed polarization intensity (\ref{expression of intensity1}) can be further rewritten as
\begin{equation}\label{expression of intensity2}
|I|=\delta^{4}l_p|\vec{B}|^{2}\sin^{2}\zeta.
\end{equation}
Finally, the observed polarized vector components along the $x$ direction and $y$ direction are
\begin{eqnarray}\label{EV01}
&&E_{x,obs}=\delta^2l_p^{\frac{1}{2}}\sin\zeta|\vec{B}|E_{x,norm}, \\ \nonumber
&&E_{y,obs}=\delta^2l_p^{\frac{1}{2}}\sin\zeta|\vec{B}|E_{y,norm},
\end{eqnarray}
and then the total polarization intensity and the electric vector position angle (EVPA) can be expressed as
\begin{eqnarray}\label{Insen01}
I=E_{x,obs}^2+E_{y,obs}^2, \quad \quad EVPA=\frac{1}{2}\arctan\frac{U}{Q}.
\end{eqnarray}
Here the Stokes parameters  $Q$ and $U$ are given by
\begin{eqnarray}\label{stroke01}
Q=E_{y,obs}^2-E_{x,obs}^2, \quad\quad\quad U=-2E_{x,obs}E_{y,obs}.
\end{eqnarray}
Combing the metric (\ref{metric}) with Eqs.(\ref{kappa0}), (\ref{EV0}), (\ref{EV01}), (\ref{Insen01}) and (\ref{stroke01}), we can study the effects of Gauss-Bonnet parameter on  polarization images of emitting ring in the 4D black hole spacetime for different magnetic fields and observing inclination angles.

\section{ Effects of Gauss-Bonnet parameter on  polarization images in the 4D  black hole spacetime}

\begin{figure}
\includegraphics[width=12cm ]{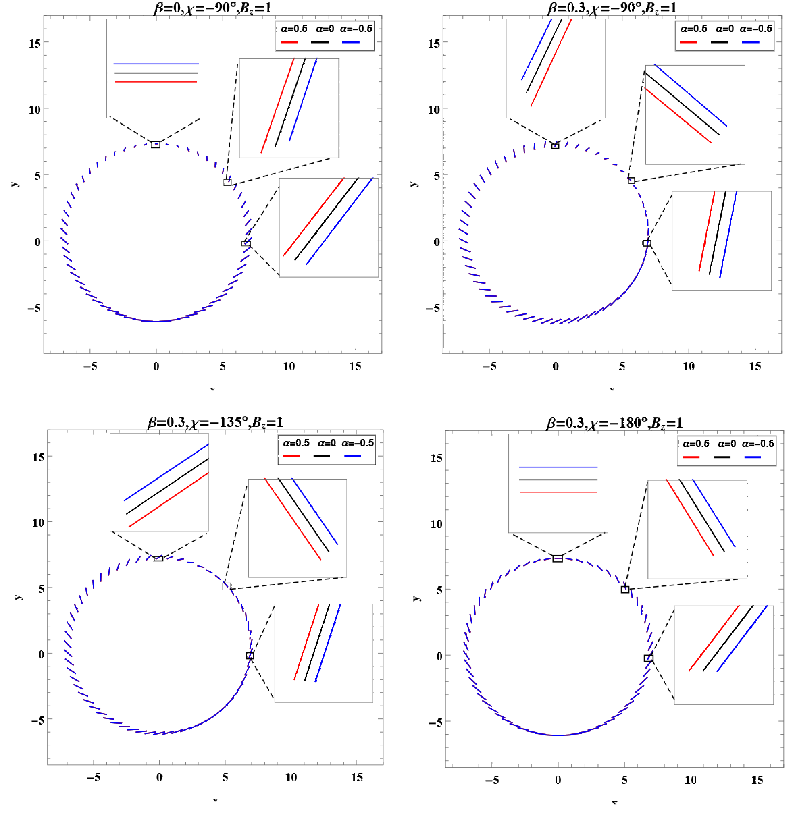}
\caption{Polarized intensity tick plots in the 4D Gauss-Bonnet black hole spacetime for different Gauss-Bonnet parameter $\alpha$ for the pure vertical magnetic field $B_z=1$. Here, we set $M=1$, the observer inclination angle $\theta=20^{\circ}$ and the ring radius $R=6$.}
\label{figip2}
\end{figure}

\begin{figure}
\includegraphics[width=12cm ]{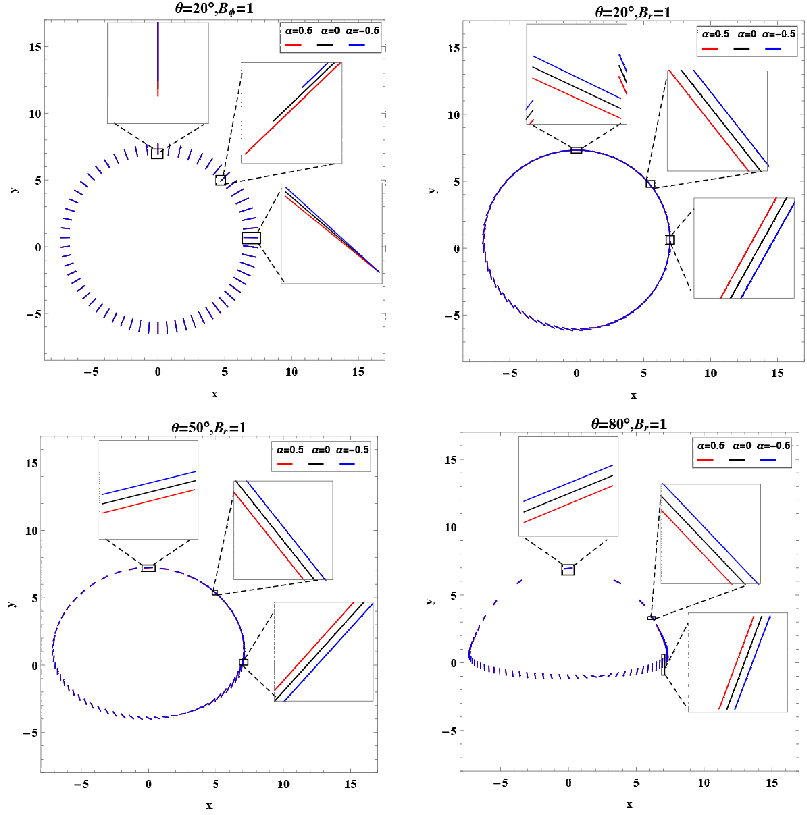}
\caption{Polarized intensity tick plots in the 4D Gauss-Bonnet black hole spacetime for different Gauss-Bonnet parameter $\alpha$ for the pure equatorial magnetic field. Here, we set $M=1$, $\beta=0.3$, $\chi=-90^{\circ}$ and $R=6$. }
\label{figip3}
\end{figure}

\begin{figure}
\includegraphics[width=14cm ]{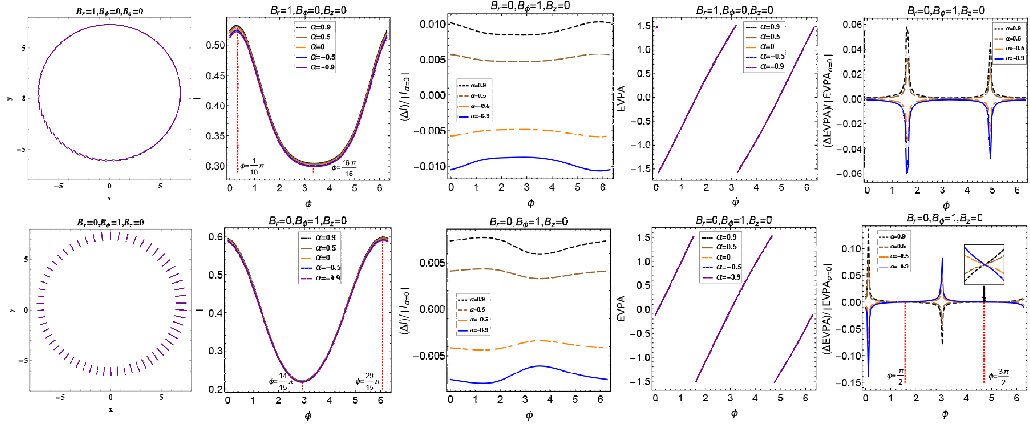}
\caption{Effects of Gauss-Bonnet parameter $\alpha$ on the polarized vector  in the 4D black hole spacetime (\ref{metric}). Here $R=6$, $\theta=20^{\circ}$, $\beta=0.3$ and $\chi=-90^{\circ}$. The top and bottom rows correspond to the cases with pure radial and pure angular magnetic fields, respectively.}
\label{figip4}
\end{figure}

\begin{figure}
\centering
\includegraphics[width=14cm]{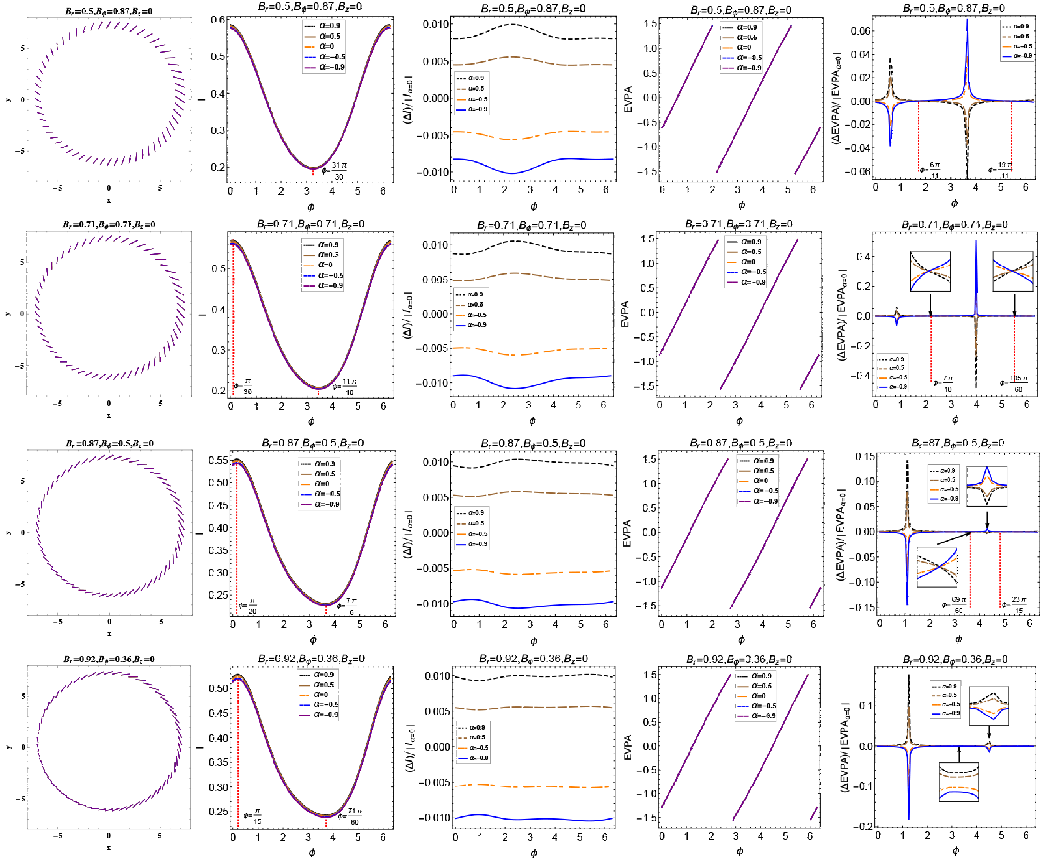}
\caption{Effects of Gauss-Bonnet parameter $\alpha$ on the polarized vector in the 4D black hole spacetime (\ref{metric}) in the cases with only the equatorial component of the magnetic field. Here $R=6$, $\theta=20^{\circ}$, $\beta=0.3$ and $\chi=-90^{\circ}$. }
\label{figip5}
\end{figure}

\begin{figure}
\includegraphics[width=14cm]{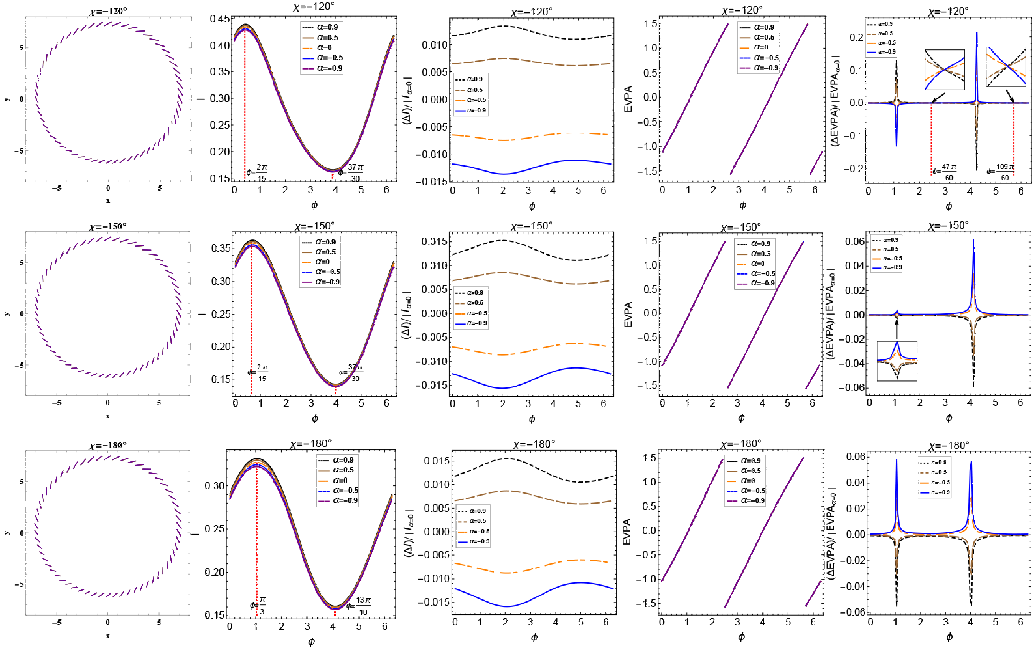}
\caption{Effects of Gauss-Bonnet parameter $\alpha$ on the polarized vector in the 4D black hole spacetime (\ref{metric}) for the different fluid direction angles $\chi$. Here $R=6$, $\beta=0.3$, $\theta=20^{\circ}$, $B_r=0.87$, $B_\phi=0.5$ and $B_z=0$. }
\label{figip6}
\end{figure}

\begin{figure}
\includegraphics[width=14cm]{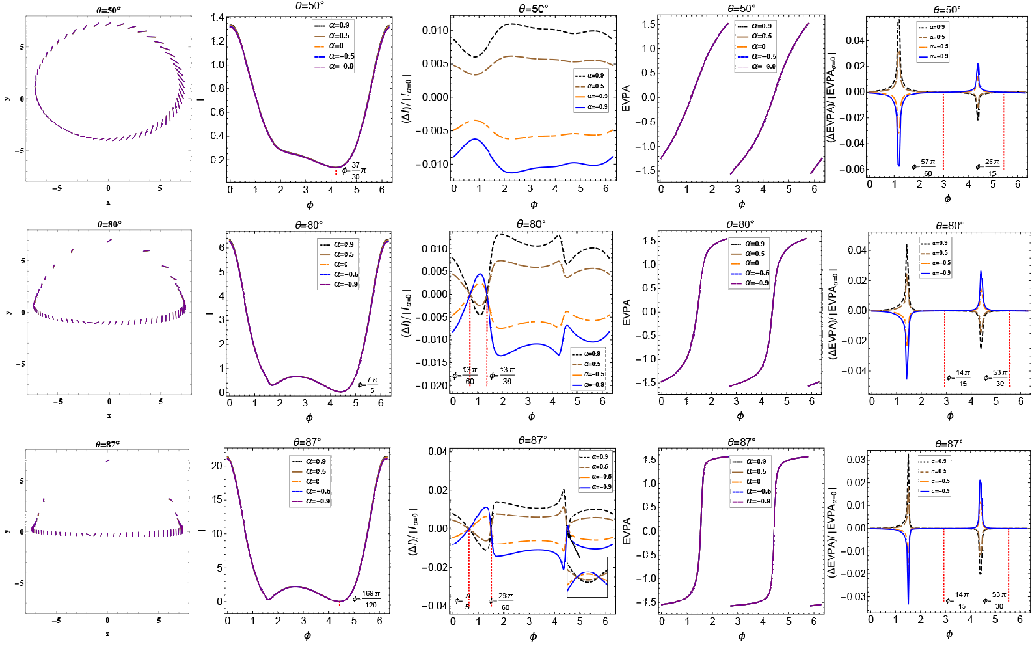}
\caption{Effects of Gauss-Bonnet parameter $\alpha$ on the polarized vector in the 4D black hole spacetime (\ref{metric}) for the different observation inclination angles $\theta$. Here $R=6$, $\beta=0.3$, $\chi=-90^{\circ}$, $B_r=0.87$, $B_\phi=0.5$ and $B_z=0$. }
\label{figip7}
\end{figure}

\begin{figure}
\includegraphics[width=14cm]{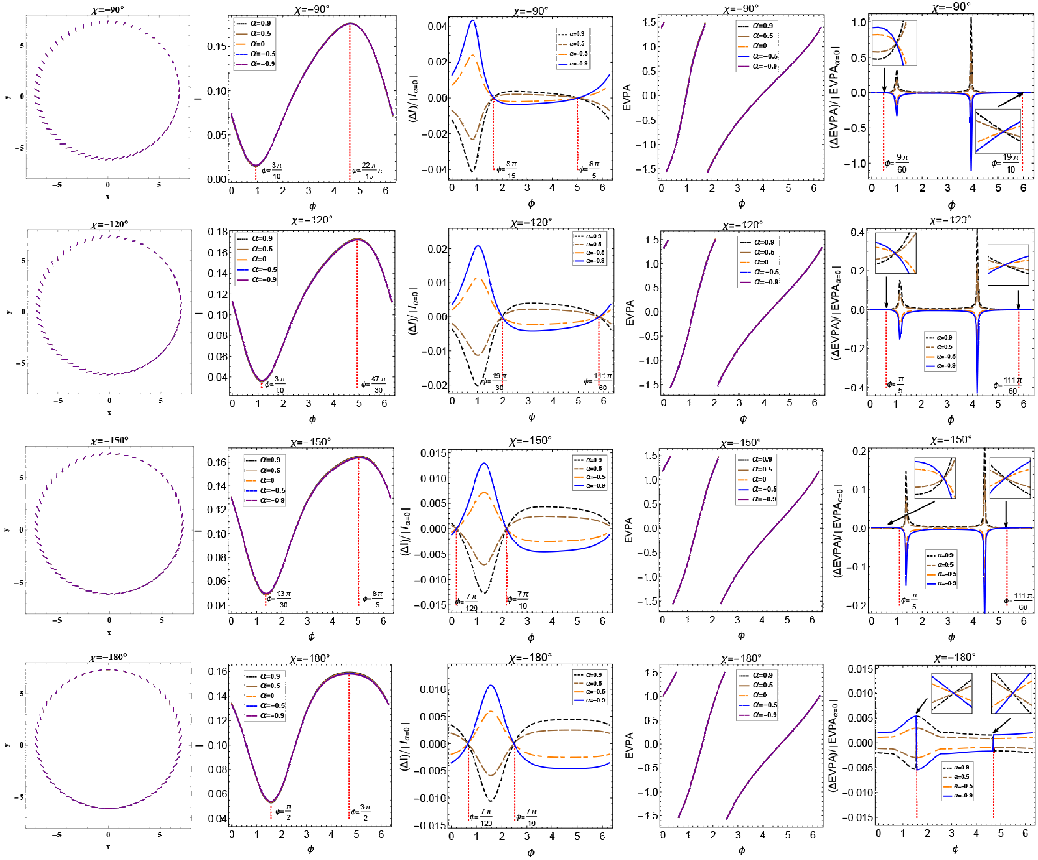}
\caption{Effects of Gauss-Bonnet parameter $\alpha$ on the polarized vector in the 4D black hole spacetime (\ref{metric}) for the different fluid direction angles $\chi$ in the case where magnetic field owns only the vertical component $B_z$.
Here $R=6$, $\beta=0.3$, $\theta=20^{\circ}$, $B_r=0$, $B_\phi=0$ and $B_z=1$.  }
\label{figip8}
\end{figure}

\begin{figure}
\includegraphics[width=14cm]{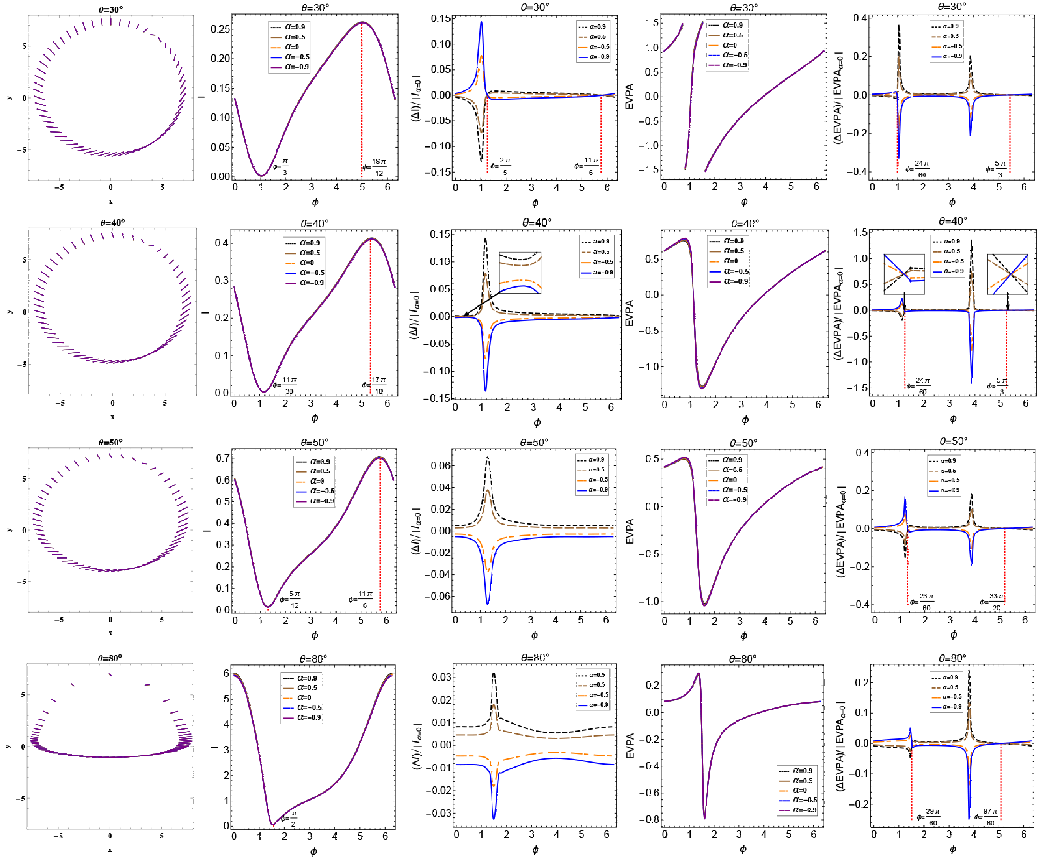}
\caption{Effects of Gauss-Bonnet parameter $\alpha$ on the polarized vector in the 4D black hole spacetime (\ref{metric}) for the different  observation inclination angles $\theta$ in the case where magnetic field owns only the vertical component $B_z$.
Here $R=6$, $\beta=0.3$, $\chi=-90^{\circ}$, $B_r=0$, $B_\phi=0$ and $B_z=1$.  }
\label{figip9}
\end{figure}

In Figs. (\ref{figip2})-(\ref{figip9}), we plot the polarization image of an equatorial emitting ring around a 4D Gauss-Bonnet black hole (\ref{metric}) for different Gauss-Bonnet parameter together with different magnetic fields, fluid velocity and observed inclination angles. Figs.(\ref{figip2}) and (\ref{figip3}) show that the dependence of polarization image on the magnetic field, the fluid velocity and the observed inclination angle in 4D Gauss-Bonnet black hole spacetime are similar to that in the usual Schwarzschild case.
Figs.(\ref{figip4})-(\ref{figip9}) present the influence of the Gauss-Bonnet coupling constant $\alpha$ on the polarization image.  In Figs.(\ref{figip4}) and (\ref{figip5}), we focus on the case with only the equatorial component of the magnetic field for fixed parameters $R=6$, $\theta=20^{\circ}$, $\beta=0.3$ and $\chi=-90^{\circ}$. For the case with only radial magnetic field $B_r$, we find that both the polarization intensity $I$ and EVPA increase with the Gauss-Bonnet parameter $\alpha$.  It is also shown in the percentage changes of polarization intensity $I$ and EVPA, where $\Delta I/I_{\alpha=0}\equiv(I_{\alpha}-I_{\alpha=0})/I_{\alpha=0}$ and $\Delta EVPA/|EVPA_{\alpha=0}|\equiv(EVPA_{\alpha}-EVPA_{\alpha=0})/|EVPA_{\alpha=0}|$. Moreover, the polarization intensity $I$ has a minimum value at the position $\phi=\frac{16}{15}\pi$ and a maximum one at the position $\phi=\frac{1}{10}\pi$.  In the case with only angular magnetic field $B_{\phi}$, the positions of the minimum and the maximum values of $I$ become $\phi=\frac{14}{15}\pi$  and  $\phi=\frac{29}{15}\pi$, respectively. With the increase of $\alpha$,
the polarization intensity $I$ increases, while the change of EVPA  depends on the position angle $\phi$.  Namely,  EVPA increases
as $0<\phi<\frac{\pi}{2}$ and $\frac{3\pi}{2}<\phi<2\pi$, and decreases as  $\frac{\pi}{2}<\phi<\frac{3\pi}{2}$. In particular, as $\phi=\frac{\pi}{2}$ or $\phi=\frac{3\pi}{2}$, one can find that EVPA is independent of the Gauss-Bonnet coupling parameter $\alpha$. In Fig.(\ref{figip5}), we present polarization images in the case where the equatorial magnetic field owns both angular and radial components. It is shown that the polarization intensity still increases with $\alpha$. The change of EVPA  becomes more complex. As $B_r<B_\phi$, one can find that the change of EVPA with $\alpha$ is similar to that in the pure angular equatorial magnetic field case. With the increasing of the radial component $B_r$, one can find that the range of $\phi$ where EVPA is a decreasing function of $\alpha$ gradually shrinks so that finally  EVPA becomes a increasing function of $\alpha$ in the total range of $\phi$ as in the case with pure radial magnetic field case.
In Fig.(\ref{figip6}), we also show the effects of Gauss-Bonnet parameter $\alpha$ on the polarized vector for different fluid
direction angles $\chi$. As the direction angle $\chi$ changes from $-120^{\circ}$ to $\chi=-180^{\circ}$, the polarization intensity still is a increasing function of $\alpha$. However, the range of $\phi$ where EVPA is a decreasing function of $\alpha$ broadens so that finally the EVPA becomes a decreasing function of $\alpha$ as $\chi=-150^{\circ}$ or $\chi=-180^{\circ}$. This means, for different Gauss-Bonnet parameter $\alpha$,  the effects of the fluid direction angle $\chi$ on the polarized vector are different from those originating from the pure radial magnetic field.
The effects of Gauss-Bonnet parameter $\alpha$ on the polarized vector for the different observation inclination angles also are shown in Fig.(\ref{figip7}) as the magnetic field lies in the equatorial plane. We observe that the peak value of the polarization intensity increases with the inclination angle. However, the change of polarized vector with Gauss-Bonnet parameter $\alpha$ becomes more complicated. As $\theta=80^{\circ}$,  the  polarization intensity is no longer a monotonically increasing function of $\alpha$. Moreover, the peak value of polarization angle deviated from that in Schwarzschild black hole spacetime increases with the observation inclination angle.
The effects of Gauss-Bonnet parameter $\alpha$ on the polarized vector in the 4D black hole spacetime (\ref{metric}) has been analyzed for the case where magnetic field owns only the vertical component $B_z$. From Fig.(\ref{figip8}), we find that the dependence of the polarization intensity and the EVPA on $\alpha$ is different from that in the case where the magnetic field the magnetic field lies in the equatorial plane. As the angle $\chi$ decreases down from $-90^{\circ}$ to $-180^{\circ}$, we find that the range of $\phi$ for polarization intensity increasing with $\alpha$ increases, but the range of $\phi$ for EVPA decreases. In Fig. (\ref{figip9}), one can find that with the increase of the inclination angle, the polarization intensity gradually becomes an increasing function of $\alpha$, but the range of $\phi$ for EVPA increasing with $\alpha$ shrinks.
\begin{figure}
\includegraphics[width=12cm]{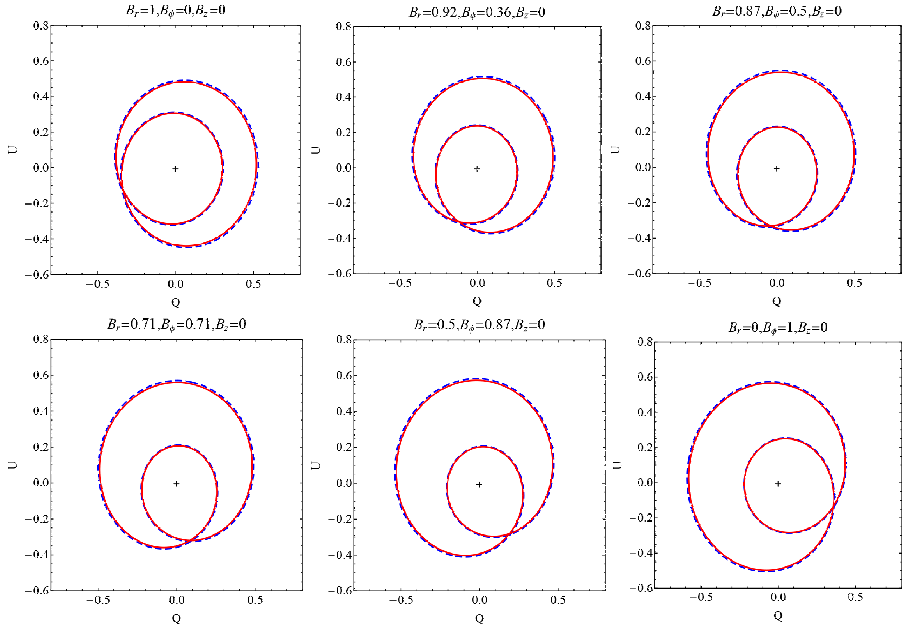}
\caption{The $Q$-$U$ diagram for different equatorial magnetic field in four-dimensional Gauss-Bonnet black hole spacetime. Here $R=6$, $\theta=20^{\circ}$, $\beta=0.3$ and $\chi=-90^{\circ}$. The blue dotted line and the red solid line represent Gauss-Bonnet constant of $\alpha=0.9$ and $\alpha=-0.9$, respectively. Black crosshairs indicate the origin of each plot.}
\label{figip10}
\end{figure}
\begin{figure}
\includegraphics[width=12cm]{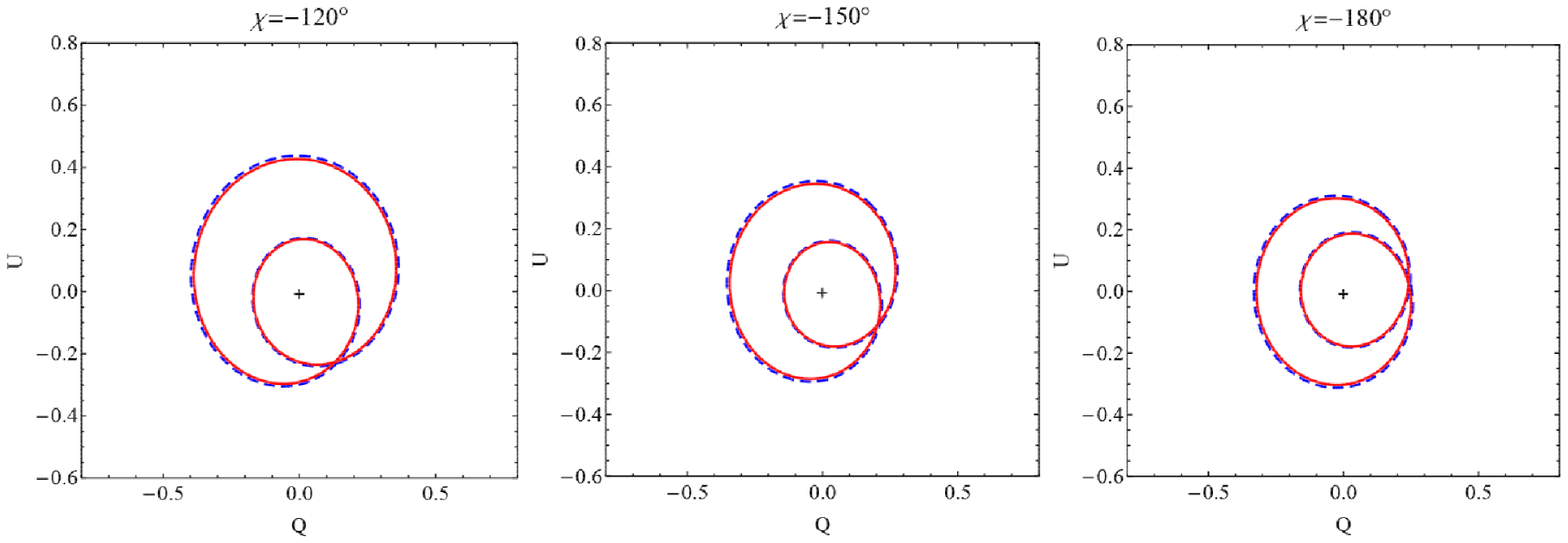}
\caption{The $Q$-$U$ diagram of equatorial  magnetic field for different fluid velocities in four-dimensional Gauss-Bonnet black hole spacetime. Here $R=6$, $\theta=20^{\circ}$, $\beta=0.3$, $B_r=0.87$, $B_\phi=0.5$ and $B_z=0$. The blue dotted line and the red solid line represent Gauss-Bonnet constant of $\alpha=0.9$ and $\alpha=-0.9$, respectively. Black crosshairs indicate the origin of each plot.}
\label{figip11}
\end{figure}
\begin{figure}
\includegraphics[width=12cm]{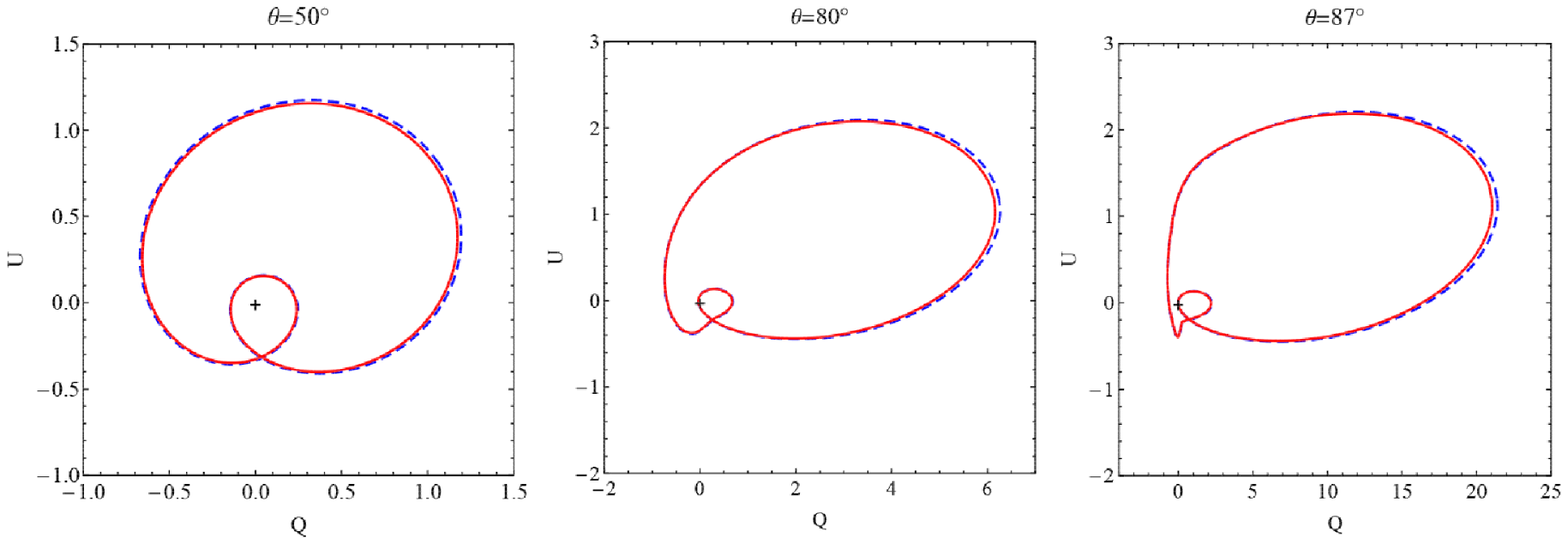}
\caption{The $Q$-$U$ diagram of equatorial  magnetic field for different observing inclination angles in four-dimensional Gauss-Bonnet black hole spacetime. Here $R=6$, $\beta=0.3$, $\chi=-90^{\circ}$, $B_r=0.87$, $B_\phi=0.5$ and $B_z=0$. The blue dotted line and the red solid line represent Gauss-Bonnet constant of $\alpha=0.9$ and $\alpha=-0.9$, respectively. Black crosshairs indicate the origin of each plot.}
\label{figip12}
\end{figure}

\begin{figure}
\includegraphics[width=12cm]{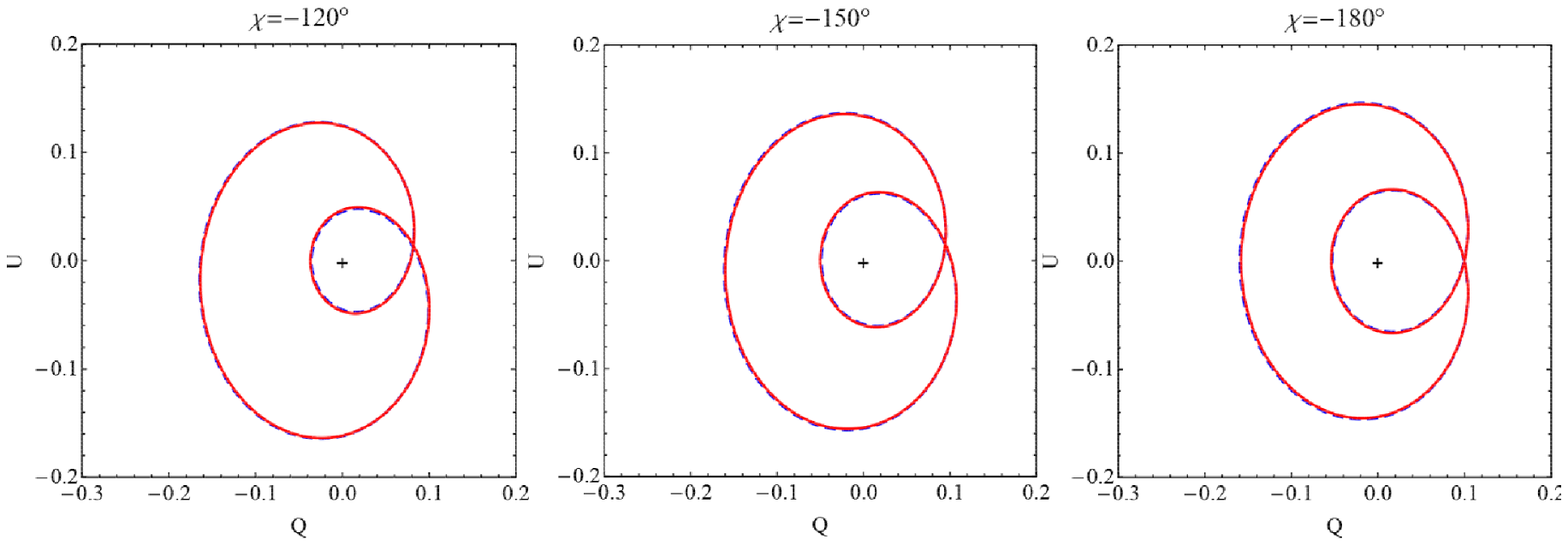}
\caption{The $Q$-$U$ diagram of pure vertical magnetic field  for different fluid velocities in four-dimensional Gauss-Bonnet black hole spacetime. Here $R=6$, $\theta=20^{\circ}$, $\beta=0.3$, $B_r=0$, $B_\phi=0$ and $B_z=1$. The blue dotted line and the red solid line represent Gauss-Bonnet constant of $\alpha=0.9$ and $\alpha=-0.9$, respectively. Black crosshairs indicate the origin of each plot.}
\label{figip13}
\end{figure}

\begin{figure}
\includegraphics[width=12cm]{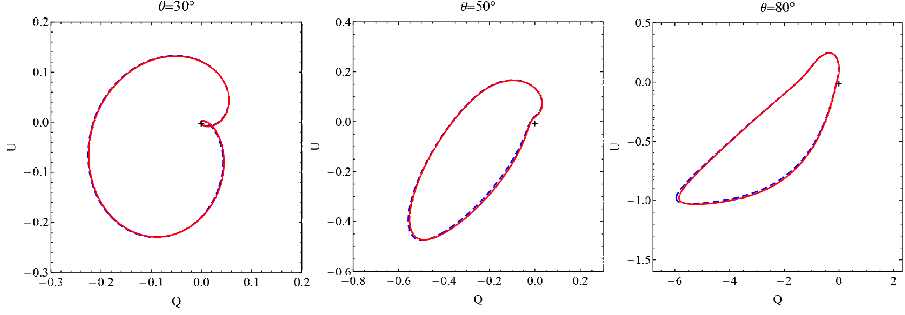}
\caption{The $Q$-$U$ diagram of pure vertical magnetic field  for different observing inclination angles in four-dimensional Gauss-Bonnet black hole spacetime. Here $R=6$, $\theta=20^{\circ}$, $\beta=0.3$, $\chi=-90^{\circ}$, $B_r=0$, $B_\phi=0$ and $B_z=1$. The blue dotted line and the red solid line represent Gauss-Bonnet constant of $\alpha=0.9$ and $\alpha=-0.9$, respectively. Black crosshairs indicate the origin of each plot.}
\label{figip14}
\end{figure}

Finally, we show effects of Gauss-Bonnet parameter $\alpha$ on the $Q$-$U$ loop diagram in the 4D black hole spacetime (\ref{metric}) in Figs. (\ref{figip10})-(\ref{figip14}). The loops in the linear Stokes $Q$-$U$ can describe the continuous variability in the polarization around a black hole. As the magnetic field lies in the equatorial plane, from Figs. (\ref{figip10})-(\ref{figip12}), one can find that there are two loops enclosing the origin in the $Q$-$U$ plane and the inner loop dramatically shrinks at high inclinations due to Doppler boosting.
As the magnetic field is vertical to the equatorial plane, from Figs. (\ref{figip13}) and (\ref{figip14}), we find that with the increase of the observer's inclination angle, the inner loop contract to a point and then vanish altogether. These features are similar to those in the Schwarzschild case. From Figs. (\ref{figip10})-(\ref{figip14}), one can find that Gauss-Bonnet constant $\alpha$  yields tiny effects in the $Q$-$U$ loops. As the observation inclination angle is smaller, the sizes of two loops in the pure equatorial magnetic field case increases with $\alpha$, but the inner loop decreases with $\alpha$ in the case of the pure vertical magnetic field. As the observation inclination angle is larger, the change of the loop size with $\alpha$ is more complicated and owns no monotonous behaviors.

\section{Summary}

We have studied the polarization image of an equatorial emitting ring around a 4D Gauss-Bonnet black hole. Our results show that the effects of Gauss-Bonnet parameter on the polarization intensity and the EVPA depend on the magnetic field configuration, the observation inclination angle, and the fluid velocity in the emitting ring. For the case with only the equatorial component of the magnetic field, as the observation inclination angle is smaller, the polarization intensity increases with Gauss-Bonnet parameter $\alpha$. As observation inclination angle $\theta$ is larger, the range of $\phi$   for the polarization intensity  increasing with $\alpha$ decreases with the inclination angle. The change of EVPA with $\alpha$ becomes more complicated. As the inclination angle $\theta=20^{\circ}$ and the fluid direction angles $\chi=-90^{\circ}$, EVPA becomes gradually an increasing function of $\alpha$ with the increase of  $B_r$. However, as the angle $\chi$ decreases, the range of $\phi$ for EVPA increasing with $\alpha$ shrinks so that EVPA becomes a decreasing function of $\alpha$ in the total range of $\phi$ as $\chi=-180^{\circ}$. Moreover, the peak value of EVPA deviated from that in Schwarzschild black hole spacetime increases with the observation inclination angle.
For the case where magnetic field owns only the vertical component $B_z$, the dependence of the polarization intensity and the EVPA on $\alpha$ is different from that in the case where the magnetic field lies in the equatorial plane. As the angle $\chi$ decreases down from $-90^{\circ}$ to $-180^{\circ}$, the range of $\phi$ for polarization intensity increasing with $\alpha$ increases, but the range of $\phi$ for the EVPA  decreases. With the increase of the inclination angle, the polarization intensity gradually becomes an increasing function of $\alpha$, but the range of $\phi$ for EVPA increasing with $\alpha$ decreases.

Finally, we also probe the effects of Gauss-Bonnet constant $\alpha$ on the $Q$-$U$ loops. As the observation inclination angle is smaller, the sizes of two loops in the pure equatorial magnetic field case increases with $\alpha$, while the inner loop decreases with $\alpha$ in the case of the pure vertical magnetic field. As the observation inclination angle is larger, the change of the loop size with $\alpha$ is more complicated and owns no monotonicity. Although the effects of Gauss-Bonnet parameter are weaker than those arising from the magnetic field,  observation inclination angle and fluid velocity, the information stored in the polarization images is beneficial to understand Gauss-Bonnet gravity.

\section{\bf Acknowledgments}

This work was  supported by the National Natural Science Foundation of China under Grant No.11875026, 11875025, 12035005, and the National Key Research and Development Program of China No. 2020YFC2201400.


\vspace*{0.2cm}

\begin{thebibliography}{99}
\baselineskip=0.5 cm
\bibitem{EHT1} The Event Horizon Telescope Collaboration, \textit{First M87 Event Horizon Telescope Results. I. The Shadow of the Supermassive Black Hole}, Astrophys. J. Lett. {\bf875}, L1 (2019).
\bibitem{EHT2} The Event Horizon Telescope Collaboration, \textit{First M87 Event Horizon Telescope Results. II. Array and Instrumentation}, Astrophys. J. Lett. {\bf875}, L2 (2019).
\bibitem{EHT3} The Event Horizon Telescope Collaboration, \textit{First M87 Event Horizon Telescope Results. III. Date Processing and Calibration}, Astrophys. J. Lett. {\bf875}, L3 (2019).
\bibitem{EHT4} The Event Horizon Telescope Collaboration, \textit{First M87 Event Horizon Telescope Results. IV. Imaging the Central Supermassive Black Hole}, Astrophys. J. Lett. {\bf875}, L4 (2019).
\bibitem{EHT5} The Event Horizon Telescope Collaboration, \textit{First M87 Event Horizon Telescope Results. V. Physical origin of the asymmetric ring}, Astrophys. J. Lett. {\bf875}, L5 (2019).
\bibitem{EHT6} The Event Horizon Telescope Collaboration, \textit{First M87 Event Horizon Telescope Results. VI. The Shadow and Mass of the Central Black Hole}, Astrophys. J. Lett. {\bf875}, L6 (2019).
\bibitem{EHT7} The Event Horizon Telescope Collaboration, \textit{First M87 Event Horizon Telescope Results. VII. Polarization of the Ring}, Astrophys. J. Lett. {\bf875}, L7 (2019).
\bibitem{EHT8} The Event Horizon Telescope Collaboration, \textit{First M87 Event Horizon Telescope Results. VIII. Magnetic Field Structure near The Event Horizon}, Astrophys. J. Lett. {\bf875}, L8 (2019).


\bibitem{PZJG1} P. A. Connors, T. Piran, and R. F. Stark, \textit{Polarization features of X-ray radiation emitted near black holes}, Astrophys. J. {\bf235}, 224-244 (1980).
\bibitem{PZJG2} B. C. Bromley, F. Melia, and S. Liu, \textit{Polarimetric Imaging of the Massive Black Hole at the Galactic Center}, Astrophys. J. Lett. {\bf555}, L83-L86 (2001).
\bibitem{PZJG3} L. Li, R. Narayan, and J. McClintock, \textit{Inferring the Inclination of a Black Hole Accretion Disk from Observations of its Polarized Continuum Radiation}, Astrophys. J. {\bf555}, 847-865 (2009).
\bibitem{PZJG4} R. V. Shcherbakov, R. F. Penna, and J. C. McKinney, \textit{Sagittarius $A^*$ Accretion Flow and Black Hole Parameters from General Relativistic Dynamical and Polarized Radiative Modeling}, Astrophys. J. {\bf755}, 847-865 (2012).
\bibitem{PZJG5} J. Dexter, \textit{A public code for general relativistic, polarised radiative transfer around spinning black holes}, Mon. Not. Roy. Astron. Soc. {\bf462}, 115-136 (2016).
\bibitem{PZJG6} R. Gold, J. C. McKinney, M. D. Johnson, and S. S. Doeleman, \textit{Probing the Magnetic Field Structure in Sgr $A^\star$ on Black Hole Horizon Scales with Polarized Radiative Transfer Simulations}, Astrophys. J. {\bf837}, 180 (2017).
\bibitem{PZJG7} F. Marin, M. Dov¡¦ciak, F. Muleri, F. F. Kislat, and H. S. Krawczynski, \textit{Predicting the X-ray polarization of type 2 Seyfert galaxies},  Mon. Not. Roy. Astron. Soc. {\bf473}, 1286-1316 (2016).
\bibitem{PZJG8} A. Jim¡äenez-Rosales and J. Dexter, \textit{The impact of Faraday effects on polarized black hole images of Sagittarius $A^\star$}, Mon. Not. Roy. Astron. Soc. {\bf478}, 1875-1883 (2018).
\bibitem{PZJG9} D. C. M. Palumbo, G. N. Wong, and B. S. Prather, \textit{Discriminating accretion states via rotational symmetry in simulated polarimetric images of m87}, Astrophys. J. {\bf894}, 156 (2020).
\bibitem{PZJG10} M. Mo¡äscibrodzka, \textit{General relativistic polarized radiative transfer with inverse-Compton scatterings}, Mon. Not. Roy. Astron. Soc. {\bf491}, 4807-4815 (2020).
\bibitem{PZJG11} M. Moscibrodzka, A. Janiuk, and M. De Laurentis, \textit{Unraveling circular polarimetric images of magnetically arrested accretion flows near event horizon of a black hole}, Mon. Not. Roy. Astron. Soc. {\bf508}, 4282-4296 (2021), arXiv: 2103.00267.
\bibitem{PZJG12} Z. Zhang, S. Chen, X. Qin and J. Jing, \textit{Polarized image of a Schwarzschild black hole with a thin accretion disk as photon couples to Weyl tensor}, Eur. Phys. J. C {\bf 81}, 991 (2021), arXiv: 2106.07981.


\bibitem{Bel} A. M. Beloborodov, \textit{Gravitational Bending of Light Near Compact Objects}, Astrophys. J. Lett. {\bf566}, L85-L88 (2002).

\bibitem{QU1} R. Abuter, A. Amorim, M. Baub$\ddot{o}$ck, J. P. Berger, H. Bonnet, W. Brandner, Y. Cl$\acute{e}$net, V. Coud$\acute{e}$ du Foresto, P. T. de Zeeuw, and et al., \textit{Detection of orbital motions near the last stable circular orbit of the massive black hole sgr A*}, Astron. Astrophys. {\bf618},  L10 (2018).
\bibitem{QU2} A. Jim$\acute{e}$nez-Rosales, J. Dexter, F. Widmann, M. Baub$\ddot{o}$ck, R. Abuter, A. Amorim, J. P. Berger, H. Bonnet, W. Brandner, and et al., \textit{Dynamically important magnetic fields near the event horizon of sgrA* }, Astron. Astrophys. {\bf643}, A56 (2020).

\bibitem{PZ1} R. Narayan, D. C. M. Palumbo, M. D. Johnson, Z. Gelles, E. Himwich, D. O. Chang, A. Ricarte, J. Dexter, C. F. Gammie, A. A. Chael, and The Event Horizon Telescope Collaboration, \textit{The Polarized Image of a Synchrotron-emitting Ring of Gas Orbiting a Black Hole}, Astrophys. J. {\bf912}, 35 (2021).
\bibitem{PZ2} Z. Gelles, E. Himwich, D. C. M. Palumbo, M. D. Johnson, \textit{Polarized Image of Equatorial Emission in the Kerr Geometry}, Phys. Rev. D {\bf 104}, 044060 (2021) arXiv: 2105.09440.

\bibitem{ringpz2} A. Jim\'{e}nez-Rosales et al., \textit{Relative depolarization of
the black hole photon ring in GRMHD models of SgrA* and M87*}, Mon. Not. Roy. Astron. Soc. {\bf503}, 4563-4575 (2021), arXiv:2103.06292 [astro-ph.HE].

\bibitem{BH} D. Glavan and C. S. Lin, \textit{Einstein-Gauss-Bonnet gravity in 4-dimensional space-time}, Phys. Rev. Lett. {\bf124}, 081301 (2020).

\bibitem{criticism1} R. A. Hennigar, D. Kubiznak, R. B. Mann and C. Pollack, \textit{On Taking the $D\rightarrow4$ limit of Gauss-Bonnet Gravity: Theory and Solutions},  J. High Energy Phys. {\bf 07}, 027 (2020), arXiv:2004.09472.
\bibitem{criticism2} S. X. Tian and Z. H. Zhu, \textit{Non-full equivalence of the four-dimensional Einstein-Gauss-Bonnet gravity and Horndeksi gravity for Bianchi type I metric}, arXiv:2004.09954.
\bibitem{criticism3} F. W. Shu, \textit{Vacua in novel 4D Einstein-Gauss-Bonnet Gravity: pathology and instability?},  Phys. Lett. B {\bf811},  135907 (2020), arXiv:2004.09339.

\bibitem{ADM} K. Aoki, M. A. Gorji and S. Mukohyama, \textit{A consistent theory of $D\rightarrow4$ Einstein-Gauss-Bonnet gravity}, Phys. Lett. B {\bf810}, 135843 (2020); arXiv:2005.03859.



\bibitem{res1} R. A. Konoplya and A. F. Zinhailo, \textit{Quasinormal modes, stability and shadows of a black hole in the novel 4D Einstein-Gauss-Bonnet gravity}, Eur. Phys. J. C. {\bf80}, 1049 (2020).
\bibitem{res2} R. A. Konoplya and A. F. Zinhailo, \textit{Grey-body factors and Hawking radiation of black holes in 4D Einstein-Gauss-Bonnet gravity}, Phys. Lett. B. {\bf810}, 145793 (2020).
\bibitem{res3} R. A. Konoplya and A. Zhidenko, \textit{$($In$)$stability of black holes in the 4D Einstein-Gauss-Bonnet and EinsteinLovelock gravities}, Phys. Dark Univ. {\bf30}, 100697 (2020).
\bibitem{res4} R. Kumar and S. G. Ghosh, \textit{Rotating black holes in 4D Einstein-GaussCBonnet gravity and its shadow}, J. Cosmol. Astropart. Phys. {\bf20}, 053 (2020).
\bibitem{res5} M.S. Churilova, \textit{Quasinormal modes of the test fields in the consistent 4D Einstein-Gauss-BonnetC(anti)de Sitter gravity}, Annals. Phys. {\bf427}, 168425 (2021).
\bibitem{res6} C.Y. Zhang, S.J. Zhang, P.C. Li, and M. Y. Guo, \textit{Superradiance and stability of the novel 4D charged EinsteinGauss-Bonnet black hole}, J. High Energy Phys. {\bf08}, 105 (2021).
\bibitem{res7} X. Qiao, L.OuYang, D. Wang, Q. Pan, and J. Jing, \textit{Holographic superconductors in 4D Einstein-GaussBonnet gravity}, J. High Energy Phys. {\bf128}, 192 (2020).
\bibitem{res8} J. Li, S. Chen, and J. Jing, \textit{Tidal effects in 4D Einstein-Gauss-Bonnet Black Hole Spacetime}, Eur. Phys. J. C. {\bf81}, 590 (2021).
\bibitem{res9} P. Liu, C. Niu, and C. Y. Zhang, \textit{Instability of the regularized 4D charged Einstein-Gauss-Bonnet de-Sitter black hole}, Chin. Phys. C. {\bf45}, 025104 (2021).
\bibitem{res10} S. Devi, R. Roy, and S. Chakrabarti, \textit{Quasinormal modes and greybody factors of the novel four dimensional Gauss-Bonnet black holes in asymptotically de Sitter space time: Scalar, Electromagnetic and Dirac perturbations}, Eur. Phys. J. C. {\bf80}, 760 (2020).
\bibitem{res11} M.Y. Guo and P. C. Li, \textit{The innermost stable circular orbit and shadow in the novel 4D Einstein-Gauss-Bonnet gravity}, Eur. Phys. J. C. {\bf80}, 588 (2020).
\bibitem{res12} S. W. Wei and Y.X. Liu, \textit{Testing the nature of Gauss-Bonnet gravity by four-dimensional rotating black hole shadow}, Eur. Phys. J. C. Plus. {\bf136}, 4 (2020).
\bibitem{res16} S. W. Wei and Y. X. Liu, \textit{Extended thermodynamics and microstructures of four-dimensional charged Gauss-Bonnet black hole in AdS space}, Phys. Rev. D. {\bf101}, 104018 (2020).
\bibitem{res13} S. J. Yang, J. J. Wan, J. Chen, J. Yang and Y. Q. Wang, \textit{Weak cosmic censorship conjecture for the novel 4D charged Einstein-Gauss-Bonnet black hole with test scalar field and particle}, Eur. Phys. J. C. {\bf80}, 937 (2020).
\bibitem{res14} Y. P. Zhang, S. W. Wei, and Y. X. Liu, \textit{Spinning test particle in four-dimensional Einstein-Gauss-Bonnet Black Hole}, Universe. {\bf6}, 103 (2020).
\bibitem{res15} M. H. Fard, H. R. Sepangi, \textit{Bending of light in novel 4D Gauss-Bonnet-de Sitter black holes by Rindler-Ishak method}, Eur. Phys. Lett.{\bf133}, 50006 (2020).
\bibitem{res17} B. E. Panah, K. Jafarzade, S. H. Hendi, \textit{Charged 4D Einstein-Gauss-Bonnet-AdS Black Holes: Shadow, Energy Emission, Deflection Angle and Heat Engine}, Nucl. Phys. B. {\bf961}, 115269 (2020).
\bibitem{res18} X. Zeng, H. Zhang, H. Zhang, \textit{Shadows and photon spheres with spherical accretions in the four-dimensional Gauss-Bonnet black hole}, Eur. Phys. J. C. {\bf80}, 872 (2020).

\bibitem{CB} C. T. Cunningham and J. M. Bardeen, \textit{The Optical Appearance of a Star Orbiting an Extreme Kerr Black Hole}, Astrophys. J. {\bf183}, 237-264 (1973).

\bibitem{PWconstant} M. Walker and R. Penrose, \textit{On quadratic first integrals of the geodesic equations for type $\{22\}$ spacetimes}, Commun.
Math. Phys. {\bf18}, 265 (2001).


\bibitem{Himwich} E. Himwich, M. D. Johnson, A. Lupsasca, and A. Strominger, \textit{Universal polarimetric signatures of the black hole photon ring}, Phys. Rev. D. {\bf101}, 084020 (2020).





\end{thebibliography}
\end{document}